\documentstyle[12pt,aps,prc,preprint]{revtex}
\newcommand{\be}{\begin{equation}}
\newcommand{\ee}{\end{equation}}
\newcommand{\ba}{\begin{eqnarray}}
\newcommand{\ea}{\end{eqnarray}}
\tightenlines
\begin{document}
\begin{titlepage}

\title{
	Limits on $\boldmath n {\bar n}$ oscillations from nuclear stability
}
\author{
	Avraham Gal \\
	Racah Institute of Physics, The Hebrew University, Jerusalem 91904,
	Israel	% insert author(s) here
}
\maketitle

\begin{abstract}
The relationship between the lower limit on the nuclear stability 
lifetime as derived from the non disappearance of 
`stable` nuclei ($T_{d}~\gtrsim~5.4~\times~10^{31}$~yr),
and the lower limit 
thus implied on the oscillation time $(\tau_{n \bar n})$ of 
a possibly underlying neutron-antineutron oscillation process, 
is clarified by studying the time evolution of the nuclear 
decay within a simple model which respects unitarity. 
The order-of-magnitude result $\tau_{n \bar n} 
\approx 2 (T_{d}/\Gamma_{\bar n})^{1/2} > 2 \times 10^{8}$ sec, 
where $\Gamma_{\bar n}$ is a typical $\bar n$ nuclear 
annihilation width, agrees as expected with the limit on
$\tau_{n \bar n}$ established by several detailed nuclear 
physics calculations, but sharply disagreeing by 15 orders 
of magnitude with a claim published recently in Phys. Rev. CRAP.
\newline \newline
$PACS$: 11.30.Fs, 13.75.Cs
\newline
{\it Keywords}: $n \bar n$ oscillations; nuclear stability
\newline \newline
%\vspace{1cm}
Corresponding author: Avraham Gal 
\newline
Tel: 972 2 6584930, Fax: 972 2 5611519
\newline
E mail: avragal@vms.huji.ac.il
\end{abstract}
\vspace{1cm}
\centerline{\today}
\end{titlepage}
\newpage

\section{Introduction}\label{Int}

The stability of nuclei, as determined by looking for
proton decay \cite{Tak86,Ber89}, places a lower limit
on the `disappearance' lifetime 
$T_{d}>(5.4 \pm 1.1) \times 10^{31}$ yr, and sets a lower 
limit also on the lifetime of other hypothetical 
processes such as neutron-antineutron ($n \bar n$) 
oscillations in free space. 
The lower limit on the free-space oscillation
time $\tau_{n \bar n}$ which has emerged from several 
quantitative nuclear physics calculations using $n$ and 
$\bar n$ optical potentials \cite{CKG81,ABM82,DGM83,HKo98} 
is approximately given by

\begin{equation}
	\tau_{n \bar n} \approx 2 (T_{d}/\Gamma_{\bar n})^{1/2}>(2.1 \pm
	0.2) \times 10^{8}~{\rm sec} \quad ,
	\label{eq:limit}
\end{equation}
where $\Gamma_{\bar n}\sim 100$ MeV is a typical $\bar n$-nuclear
annihilation width. This slightly exceeds the limit
\begin{equation}
	\tau_{n \bar n} > 0.9 \times 10^{8}~{\rm sec}
	\label{eq:tau}
\end{equation}
set in the ILL-Grenoble reactor experiment \cite{Bal94}. A new
experiment planned at ORNL will hopefully improve this limit by two
orders of magnitude \cite{Kam97}.

Eq. (\ref{eq:limit}) may be rewritten as

\begin{equation}
	T_{d} \approx {1 \over 4} (\Gamma_{\bar n}\tau_{n \bar n}) \tau_{n
	\bar n} \quad ,
	\label{Td1}
\end{equation}
showing that, since $\Gamma_{\bar n}\tau_{n \bar n} \gg 1$ owing to
the huge lifetime distinction between strong interactions (lifetime
$\sim \Gamma_{\bar n}^{-1}$) and superweak interactions (lifetime
$\sim \tau_{n \bar n}$), the nuclear disappearance lifetime $T_{d}$
induced by $n \bar n$ oscillations is many orders of magnitude longer
than the $n \bar n$ oscillation lifetime in free space. This is
equivalent to the common statement that $n \bar n$ oscillations in
matter undergo severe suppression, since the neutron and the
antineutron feel nuclear potentials which are extremely different from
each other, and the mass degeneracy which allows such pure
oscillations between a free $n$ and a free $\bar n$ is thus removed in
the nuclear medium.

Nazaruk \cite{Naz94,Naz98} has raised objections to the use of
nuclear physics potential models, claiming that $n \bar n$
oscillations are {\it not} suppressed at all in the nuclear medium.
Thus, in the first paper \cite{Naz94} he obtained $T_{d} \sim \tau_{n
\bar n}$, resulting in the limit $\tau_{n \bar n} > 10^{31}$ yr which
advances the lower limit (\ref{eq:limit}) by about 30 orders of
magnitude! Dover, Gal, Richard \cite{DGM95} and Krivoruchenko 
\cite{Kri96} subsequently pinpointed errors in that paper and 
reestablished the generally accepted lower limit (\ref{eq:limit}).
In the second, more recent paper \cite{Naz98}, Nazaruk `rederived'
the $T_{d} \sim \tau_{n \bar n}$ result, but argued that since the
nuclear decay is non exponential, a more careful consideration of the
nuclear stability limit translates into $\tau_{n \bar n} >
10^{16}$ yr, which advances the lower limit (\ref{eq:limit}) 
by `only' 15 orders of magnitude. Such far-reaching claims, when published
by a first-rate journal, should not go unanswered.

In this note I wish to expand the arguments outlined by Dover {\it et
al.} \cite{DGM95}, in order to show in detail how 
Eq. (\ref{eq:limit}) for $\tau_{n \bar n}$ is obtained. 
In Sec. \ref{Osc} we consider the $n \bar n$ mass matrix in $\bar
n$-annihilating matter, in order to derive in the most economical way
the eigen-lifetimes in a toy seesaw model. In Sec. \ref{Tdep} we
study in detail the time evolution of the $n \bar n$ `oscillating'
system in matter. In the concluding
Sec. \ref{Disc} we have tried to identify the error in Nazaruk's logic.

\section{Oscillations --- the seesaw mechanism}\label{Osc}

A common approach in problems of oscillations, be it for CP symmetry
or in neutrino mass generation problems, is to diagonalize the mass
matrix in order to find out the physical states. For $n \bar n$
oscillations induced by a coupling $\varepsilon$, the simplest form
of the in-medium mass matrix is

\begin{equation}
\left( {\matrix{m&\varepsilon \cr
\varepsilon &{m-i{{\Gamma _{\bar n}} \over 2}}\cr
}} \right)
	\label{eq:matrix}
\end{equation}
where $m$ is the joint value of the $n$ and $\bar n$ masses, and
$\Gamma_{\bar n} \sim 100$ MeV is the annihilation width of the $\bar
n$ in the nuclear medium. For $\Gamma _{\bar n} = 0$, i.e. in free
space, the two eigenmasses are $m \pm \varepsilon$, differing only
by a tiny $\varepsilon$ from each other. The reality of these
eigenmasses gives rise to a purely oscillatory behavior $n \leftrightarrow
\bar n$, with lifetime $\tau_{n \bar n} = \varepsilon ^{-1}$. In the
nuclear medium, when $\Gamma _{\bar n} \not= 0$, the eigenmasses of
(\ref{eq:matrix}) are given by

\begin{equation}
m_{eigen}=m-{\textstyle{1 \over 4}}i\Gamma _{\bar n}\pm {\textstyle{1 \over 4}}
i\Gamma _{\bar n}\left( {1-{{16\varepsilon ^2} \over {\Gamma _{\bar n}^2}}}
\right)^{{\raise3pt\hbox{$\scriptstyle 1$} \!\mathord{\left/
{\vphantom {\scriptstyle {1 2}}}\right.\kern-\nulldelimiterspace}
\!\lower3pt\hbox{$\scriptstyle 2$}}} \ \ ,
	\label{eq:eigmass1}
\end{equation}
which to leading order in $\varepsilon^{2}/\Gamma_{\bar n}^{2}$ assume
the values

\begin{equation}
m_n=m-{1 \over 2}i{{4\varepsilon ^2} \over {\Gamma _{\bar n}}}+\ldots \    ,\
m_{\bar n}=m-{1 \over 2}i\Gamma _{\bar n}+\ldots \    ,
	\label{eq:eigmass2}
\end{equation}
where the dots stand for higher order terms. The degeneracy of the
real parts of $m_{eigen}$ can easily be removed by introducing
$n$-nucleus and $\bar n$-nucleus real optical potentials
 which differ substantially from each other. 
In discussing decay
modes, however, only Im $m_{eigen}$ matters; adding real potentials
does not change the splitting between, and the order of magnitude of
the two values for Im $m_{eigen}$ in Eq. (\ref{eq:eigmass2}).
Therefore, for the sake of simplicity, we ignore these real potentials.
Identifying the eigenwidth $\gamma$ with $-2$ Im $m_{eigen}$, one
obtains from Eq. (\ref{eq:eigmass2}) two eigenwidths:

\begin{equation}
	\gamma_{n}={{4\varepsilon}\over{\Gamma_{\bar n}}}
	\varepsilon \ \ \ , \ \ \ \gamma_{\bar n} = \Gamma_{\bar n} \ \ .
	\label{eq:eigwidth1}
\end{equation}
Intuitively, since $\gamma_{\bar n} = \Gamma_{\bar n}$ is the nuclear
annihilation width of an antineutron, $\gamma_{n}$ must stand for the
decay rate of the neutron (in units where $\hbar = 1$). It is clear
that this decay rate undergoes a huge suppression factor, $4
\varepsilon / \Gamma _{\bar n} \ll 1$, with respect to the free-space
oscillation rate $\varepsilon$. The huge disparity between
$\gamma_{n}$ and $\gamma_{\bar n}$ is a good demonstration of the
{\it seesaw mechanism} encountered in the discussion of neutrino mass
generation problems; note that the product $\gamma_{n} \gamma_{\bar
n} = 4 \varepsilon^{2}$ is of the same order of magnitude as in free
space, which is a necessary condition for the seesaw mechanism. These
qualitative arguments need to be explored quantitatively by studying
the time evolution of $n \bar n$ `oscillations' in the nuclear
medium. This is done in the next section.

\section{Explicit time evolution}\label{Tdep}

For simplicity, following Ref. \cite{DGM95}, we write the time
dependent coupled Schr\"odinger equations for zero momentum neutron
and antineutron in nuclear matter $(\hbar = 1)$:

\begin{equation}
i\partial_{t}\psi _n=\varepsilon \psi _{\bar n}\  ,\
i\partial_{t}\psi _{\bar n}=\varepsilon \psi
_n-i{{\Gamma _{\bar n}} \over 2}\psi _{\bar n}\  .
	\label{eq:coupled}
\end{equation}
No nuclear (real) potentials $U_n$ and $U_{\bar n}$ 
appear in the present discussion, which is focussed 
on the interplay and competition between the free-space 
oscillation rate $\varepsilon = \tau _{n \bar n}^{-1}$ 
and the $\bar n$ decay rate $\Gamma_{\bar n}$ due to the 
strong-interaction $\bar n$ nuclear annihilation, 
and their effect
will be briefly discussed later on. The coupled first order
equations (\ref{eq:coupled}) give rise to the following 
second order differential equation for each one of 
$\psi _{n}, \psi _{\bar n}$:

\begin{equation}
\left( {\partial^{2}_{t}+{1 \over
2}\Gamma _{\bar n}\partial_{t}+\varepsilon ^2} \right)\psi
=0\   .
	\label{eq:second}
\end{equation}
Seeking eigensolutions of the form 
$\psi_{j}=\exp(i\omega_{j}t/2)$, the
eigenfrequencies $\omega_{j}$ satisfy a quadratic equation

\begin{equation}
\omega ^2-i\Gamma _{\bar n}\omega -4\varepsilon ^2=0\   ,
	\label{eq:eigfreq1}
\end{equation}
yielding two solutions:

\begin{equation}
\omega _{\bar n}=i\Gamma _{\bar n}\left\{ {1-{1 \over 2}\left[
{1 -\left( {1-16{{\varepsilon ^2} \over {\Gamma _{\bar n}^2}}}
\right)^{{\raise3pt\hbox{$\scriptstyle 1$} \!\mathord{\left/
{\vphantom {\scriptstyle {1 2}}}\right.\kern-\nulldelimiterspace}
\!\lower3pt\hbox{$\scriptstyle 2$}}}} \right]} \right\}=i\Gamma _{\bar n}
\left( {1-4{{\varepsilon ^2} \over {\Gamma _{\bar n}^2}}+\ldots } \right)\   ,
	\label{eq:eigfreq2}
\end{equation}

\begin{equation}
\omega _n=i\Gamma _{\bar n}{1 \over 2}\left[ {1-\left( {1-16{{\varepsilon ^2}
\over {\Gamma _{\bar n}^2}}} \right)^{{\raise3pt\hbox{$\scriptstyle 1$}
\!\mathord{\left/ {\vphantom {\scriptstyle {1 2}}}
\right.\kern-\nulldelimiterspace} \!\lower3pt\hbox{$\scriptstyle 2$}}}}
\right]=i{{4\varepsilon ^2} \over {\Gamma _{\bar n}}}\left(
{1+4{{\varepsilon ^2} \over {\Gamma _{\bar n}^2}}+\ldots } \right)\   ,
	\label{eq:eigfreq3}
\end{equation}
where the expansion in even power of the extremely small quantity
$\varepsilon ^{2}/\Gamma_{\bar n}^{2} \ll 1$ is indicated. These
eigenfrequencies are related to the eigenwidths in Eq. (\ref{eq:eigwidth1})
by $\omega = i \gamma$. The eigenfrequency $\omega _{\bar n}$ gives
rise to exponential decay of $| \psi |^{2}$ with a rate $\Gamma
_{\bar n}$, to leading order. Dover {\it et al.} \cite{DGM95} noted
that $\omega_{\bar n}$ should be discarded since it corresponds to the
rate of disappearance of an antineutron when its oscillation
coupling $\varepsilon$ to the neutron is neglected. It was further
noted that the eigenfrequency $\omega _{n}$ gives rise to a rate of
disappearance $4 (\varepsilon / \Gamma_{\bar n}) \varepsilon$ which
was then interpreted as that for a neutron, due to its oscillation
coupling $\varepsilon$ to the antineutron. This latter rate is
equivalent to the estimate given in Eq. (\ref{Td1}) for $T_{d}$. We
note that since $\omega _{n} \omega _{\bar n}=-4 \varepsilon ^{2}$,
the huge disparity between the frequencies $\omega _{n}$ and $\omega
_{\bar n}$ is a good demonstration of the seesaw effect encountered in
the discussion of scales of masses in high energy physics.

Here we wish to proceed in greater detail and rigor, using the
eigenfrequencies (\ref{eq:eigfreq2}, \ref{eq:eigfreq3}), to construct
the wavefunctions $\psi _j$ with the appropriate {\it temporal boundary
conditions} for the neutron ($j=n$) and for the antineutron ($j=\bar n$). 
In considering the nuclear disappearance lifetime $T_{d}$, 
these boundary conditions are

\begin{equation}
	\psi_{n}(t = 0) = 1 \ , \quad \psi _{\bar n} (t = 0) = 0 \ \ .
	\label{eq:0time}
\end{equation}
The corresponding linear combinations of the eigensolutions 
are then given by

\begin{equation}
\psi _{n}(t)={{\omega _{\bar n}} \over {\omega _{\bar n}-\omega _n}}
\exp \left( {{{i\omega _nt} \mathord{\left/ {\vphantom {{i\omega _nt} 2}}
\right. \kern-\nulldelimiterspace} 2}} \right)-{{\omega _n} \over
{\omega _{\bar n}-\omega _n}}\exp \left( {{{i\omega _{\bar n}t}
\mathord{\left/ {\vphantom {{i\omega _{\bar n}t} 2}} \right.
\kern-\nulldelimiterspace} 2}} \right) \ ,
	\label{eq:psi1}
\end{equation}

\begin{equation}
\psi _{\bar n}(t)={{2\varepsilon } \over {\omega _{\bar n}-\omega _n}}\left(
{\exp \left( {{{i\omega _nt} \mathord{\left/ {\vphantom {{i\omega _nt} 2}}
\right. \kern-\nulldelimiterspace} 2}} \right)-\exp \left( {{{i\omega _{\bar n}t}
\mathord{\left/ {\vphantom {{i\omega _{\bar n}t} 2}} \right.
\kern-\nulldelimiterspace} 2}} \right)} \right)\   .
	\label{eq:psi2}
\end{equation}
Note that since $\omega_{n}/\omega_{\bar n} \approx 4
\varepsilon^{2}/\Gamma _{\bar n}^{2} \ll 1$, $\psi_{n}(t)$ is dominated
at all times by the eigensolution with eigenfrequency $\omega _{n}$.
It is also clear, by inspecting Eq. (\ref{eq:eigfreq2}), that the
$\bar n$ wavefunction which evolves from zero at $t = 0$ always
remains very small with respect to the neutron wavefunction:

\begin{equation}
\left| {{{\psi _{\bar n}(t)} \mathord{\left/ {\vphantom 
{{\psi _{\bar n}(t)}
{\psi _n}(t)}} \right. \kern-\nulldelimiterspace} {\psi _n}(t)}}
\right|^2 \sim {{4\varepsilon ^2} \over {\Gamma _{\bar n}^2}}\  ,
	\label{eq:ratio}
\end{equation}
this approximate relationship becoming exact for times $t \gg
\Gamma_{\bar n}^{-1}$ when the second exponent on the right-hand side
of Eqs. (\ref{eq:psi1}, \ref{eq:psi2}) 
may be safely dropped out. For such `long' times, 
the time dependence of both $\psi_{n}$ and
$\psi_{\bar n}$ is given by

\begin{equation}
\left| \psi (t) \right|^2 \sim \exp \left( {-4{{\varepsilon ^2}
\over {\Gamma _{\bar n}}}t} \right)\  ,
	\label{eq:dec1}
\end{equation}
so that the rate of disappearance $\Gamma_{d}$ may be read off this
exponential decay:

\begin{equation}
	\Gamma_{d}={{4\varepsilon ^2} \over {\Gamma _{\bar n}}} \ \ .
	\label{eq:width}
\end{equation}
Recalling that $\varepsilon = \tau^{-1}_{n \bar n}$, we obtain for
the disappearance lifetime

\begin{equation}
T_d=\Gamma _d^{-1}={1 \over 4}\left( {\Gamma _{\bar n}\tau _{n\bar n}}
\right)\tau _{n\bar n}\  ,
	\label{eq:Td2}
\end{equation}
which agrees precisely with Eq. (\ref{Td1}) of Sec. \ref{Int}.

The minimal model of Eq. (\ref{eq:coupled}) for the time evolution
of $\psi_{n}$ and $\psi_{\bar n}$ may be extended to include real
potentials $U_{n}$ and $U_{\bar n}$, and spatial structure:

\begin{equation}
i\partial_{t}\psi _n=-{\Delta  \over {2m}}\psi _n+U_n(r)\psi
_n+\varepsilon \psi _{\bar n}\   ,
	\label{eq:schr1}
\end{equation}

\begin{equation}
i\partial_{t}\psi _{\bar n}=-{\Delta  \over {2m}}\psi
_{\bar n}+\left( {U_{\bar n}(r)-iW(r)} \right)\psi _{\bar n}+\varepsilon \psi _n\   ,
	\label{eq:schr2}
\end{equation}
with the same $t=0$ boundary conditions specified in Eq.
(\ref{eq:0time}). In order to obtain the decay rate of the neutron,
we multiply Eq. (\ref{eq:schr1}) for $\psi_{n}$ by $\psi_{n}^{*}$, and
the corresponding equation for $\psi_{n}^{*}$ by $\psi_{n}$, subtract
from each other and integrate over space. The result is

\begin{equation}
-\partial_{t}\int {\left| {\psi _n} \right|^2d^3r} =
2\varepsilon \int {{\rm Im} \left( {\psi _{\bar n}^*\psi _n} \right)d^3r\   .}
	\label{eq:rate1}
\end{equation}
This decay rate vanishes for $\varepsilon \to 0$, but the precise
power of $\varepsilon$ by which it occurs depends on the ratio
$|\psi_{\bar n}/\psi_{n}|$. Acting similarly on Eq. (\ref{eq:schr2})
for $\psi_{\bar n}$, the decay rate of the antineutron is obtained:

\begin{equation}
-\partial_{t}\int {\left| {\psi _{\bar n}} \right|^2d^3r}=\int
{2W\left| {\psi _{\bar n}} \right|^2d^3r}-2\varepsilon \int {{\rm Im} \left(
{\psi _{\bar n}^*\psi _n} \right)d^3r\   .}
	\label{eq:rate2}
\end{equation}
We note that the $2 \varepsilon \int$ terms in Eqs. (\ref{eq:rate1},
\ref{eq:rate2}) appear with opposite signs; what's lost out of the
neutron intensity due to the oscillation coupling $\varepsilon$ is
precisely gained by the $\bar n$ intensity, and vice versa, on top of
the $\bar n$ annihilation decay rate. The {\it total} decay rate is
obtained by adding up Eqs. (\ref{eq:rate1}, \ref{eq:rate2}):

\begin{equation}
-\partial_{t}\int {\left( {\left| {\psi _n}
\right|^2+\left| {\psi _{\bar n}} \right|^2} \right)d^3r}=\int
{2W\left| {\psi _{\bar n}} \right|^2d^3r\   ,}
	\label{eq:rate3}
\end{equation}
showing that if the antineutron did not annihilate in the nuclear
medium $(W=0)$, there would be no loss of intensity from the combined
$n, \bar n$ space. A rough estimate of $|\psi_{\bar
n}|^{2}/|\psi_{n}|^{2}$ may be made by inspecting Eq. (\ref{eq:schr2})
and noticing that, for the boundary conditions Eq. (\ref{eq:0time}),
the evolved $(U_{\bar n}-iW)\psi_{\bar n}$ term on the right-hand side 
of Eq. (\ref{eq:schr2}) should be of the same order of magnitude as the
source term $\varepsilon \psi _{n}$ which generates it. Since
$U_{\bar n}$ and $W \sim \Gamma_{\bar n}/2$ are all of the same order
of magnitude, we obtain

\begin{equation}
{{\left| {\psi _{\bar n}} \right|^2} \over {\left| {\psi _n}
\right|^2}}\sim{{\varepsilon ^2} \over {W^2}}\sim{{4\varepsilon ^2}
\over {\Gamma _{\bar n}^2}}\   ,
	\label{eq:est1}
\end{equation}
which agrees with Eq. (\ref{eq:ratio}). The properly normalized decay
rate $\gamma$ is then given, using Eq. (\ref{eq:rate3}), by

\begin{equation}
\gamma \equiv {{-\partial_{t}\int {\left( {\left| {\psi _n}
\right|^2+\left| {\psi _{\bar n}} \right|^2} \right)d^3r}} \over {\int {\left(
{\left| {\psi _n} \right|^2+\left| {\psi _{\bar n}} \right|^2} \right)d^3r}}}=
{{\int {2W\left| {\psi _{\bar n}} \right|^2d^3r}} \over {\int {\left( {\left|
{\psi _n} \right|^2+\left| {\psi _{\bar n}} \right|^2} \right)d^3r}}}\   ,
	\label{eq:ratef}
\end{equation}
which is approximately of order

\begin{equation}
\gamma \sim 2W\left| {{{\psi _{\bar n}} \over {\psi _n}}} \right|^2
\sim{{4 \varepsilon ^2} \over {\Gamma _{\bar n}}}\   ,
	\label{eq:est2}
\end{equation}
agreeing with the order of magnitude of $\Gamma_{d}$, Eq.
(\ref{eq:width}). We note that Eq. (\ref{eq:ratef}) expresses the loss
of $n$ and $\bar n$ intensities to the unspecified final nuclear
debris products.

\section{Discussion}\label{Disc}

We outlined in the previous sections, using a simple and transparent
potential model, how the free-space $n \bar n$ oscillation period
$\tau_{n \bar n}$ gets tremendously prolonged in the nuclear medium,
by a factor $\Gamma_{\bar n} \tau_{n \bar n} /4$, to yield the
corresponding nuclear decay lifetime $T_{d}$ of Eq. (\ref{eq:Td2}).
More detailed calculations \cite{ABM82,DGM83}, which treat the nuclear
medium as a dynamical entity, confirm this order-of-magnitude estimate
of $T_{d}$. It is then perplexing to encounter occasionally claims
that $n \bar n$ oscillations are {\it not} suppressed in the nuclear
medium. Below we counter the most recent claim of this sort made by
Nazaruk \cite{Naz98}.

Nazaruk argued that potential models involve double counting by
allowing for $\bar n$-nucleus elastic and inelastic scattering, and
that this is manifested within such models by a calculated nuclear 
decay probability which is linear in the time $t$ instead of the 
(correctly anticipated) $t^{2}$ dependence for `short' 
oscillation times. Our response is
that the only nuclear property of the antineutron included in the
present potential model is its nuclear decay width $\Gamma_{\bar n}$,
unrelated in this model to any underlying $\bar n$ scattering process.
Furthermore, we do recover a leading $t^{2}$ dependence by expanding
Eqs. (\ref{eq:psi1}, \ref{eq:psi2}) in powers of time $t$:

\begin{equation}
1-\left| {\psi _n} \right|^2=\varepsilon ^2t^2-{1 \over 6}\varepsilon
^2\Gamma _{\bar n}t^3+\ldots \   ,
	\label{eq:dep1}
\end{equation}

\begin{equation}
\left| {\psi _{\bar n}} \right|^2=\varepsilon ^2t^2-{1 \over 2}\varepsilon
^2\Gamma _{\bar n}t^3+\ldots \   ,
	\label{eq:dep2}
\end{equation}
so that the neutron depletion rate would appear to be $\varepsilon$,
as in free space --- unaffected by the nuclear medium, in particular
by the $\bar n$ annihilation width $\Gamma_{\bar n}$, and agreeing
with Nazaruk's thesis. However, this expansion is useful only for
extremely short times, such that $\Gamma_{\bar n}t \ll 1$, before
the $\bar n$ annihilation can have its act and exercise its toll. For
relevant times which satisfy $\Gamma_{\bar n}t \gg 1$, the expansion
(\ref{eq:dep1}, \ref{eq:dep2}) becomes useless. As argued here in Sec.
\ref{Tdep}, however, for such (still short) times, the $\omega_{\bar
n}$ exponent in Eqs. (\ref{eq:psi1}, \ref{eq:psi2}) may safely be
neglected, resulting in the following exponential decays

\begin{equation}
\left| {\psi _n\left( t \right)} \right|^2\to \exp \left( {-4{{\varepsilon ^2}
\over {\Gamma _{\bar n}}}t} \right)\  ,\    \left| {\psi _{\bar n}\left( t
\right)} \right|^2\to {{4\varepsilon ^2} \over {\Gamma _{\bar n}^2}}\exp
\left( {-4{{\varepsilon ^2} \over {\Gamma _{\bar n}}}t} \right)\  ,\  \
	\label{eq:dec2}
\end{equation}
with a rate of disappearance $\Gamma_{d}= 4{{\varepsilon^{2}}/{\Gamma
_{\bar n}}}$ (Eq. (\ref{eq:width})) which is directly read off
this exponential decay.

The trouble with Nazaruk's arguments may be traced as follows.
Using S-matrix manipulations, for $t \gg \Gamma_{\bar n}^{-1}$ he
reaches the expression

\begin{equation}
	w(t) \approx \varepsilon^{2}t^{2}w_{\bar n}(t)
	\label{eq:prob1}
\end{equation}
for the `probability' $w(t)$ of nuclear decay in terms of the $\bar
n$ decay probability $w_{\bar n}(t)$ (see Eqs. (\ref{eq:eigfreq2},
\ref{eq:width}, \ref{eq:Td2}) of Ref. \cite{Naz98}). He then assumes
that, since over this time all the antineutrons 
got already annihilated, 
$w_{\bar n}(t) \approx 1$. 
This is wrong: the $\bar n$ decay probability is a conditional one, 
depending on how many antineutrons were there in the
first place with respect to neutrons. Unitarity prevents the
accumulation of too many $\bar n$, and as seen from Eq. (\ref{eq:dec2})

\begin{equation}
w_{\bar n} \left( {\Gamma _{\bar n}^{-1} \ll t \ll {{\Gamma _{\bar n}}
\mathord{\left/ {\vphantom {{\Gamma _{\bar n}} {\varepsilon ^2}}} \right.
\kern-\nulldelimiterspace} {\varepsilon ^2}}} \right)\approx {{4\varepsilon ^2}
\over {\Gamma _{\bar n}^2}}\   .
	\label{eq:prob2}
\end{equation}
Hence

\begin{equation}
w\left( t \right) \approx {{4\varepsilon ^4} \over {\Gamma _{\bar n}^2}}t^2\   ,
	\label{eq:prob3}
\end{equation}
so that the neutron disappearance rate is of order $2
{\varepsilon^{2}}/{\Gamma_{\bar n}}$, in agreement with 
Eq. (\ref{eq:est2}) and with our result
Eq. (\ref{eq:width}) for $\Gamma_{d}$.

To summarize, we have confirmed the well known lower limit obtained on $n
\bar n$ oscillations in free space from the stability of nuclei, as
given by the estimate of Eq. (\ref{eq:limit}) which states that
$\tau_{n \bar n} \gtrsim 2 \times 10^{8}$ sec. This estimate agrees
to within better than a factor of two with the series of quantitative
calculations by Alberico {\it et al.} \cite{ABM82} and Dover {\it et
al.} \cite{DGM83}. There is good reason then to plan reactor
experiments \cite{Kam97} which promise to push up this lower limit by
perhaps as much as two orders of magnitude.

\vspace{15 mm}

This research was partially supported by the Israel Science Foundation


\begin{thebibliography}{99}
 	\bibitem{Tak86}  M. Takita {\it et al.} (Kamiokande collaboration),
 	Phys. Rev. D {\bf 34}, 902 (1986).

 	\bibitem{Ber89}  Ch. Berger {\it et al.} (Fr\`ejus collaboration),
 	Phys. Lett. B {\bf 240}, 237 (1989).

 	\bibitem{CKG81}  K.G. Chetyrkin, M.V. Kazarnovsky, V.A. Kuzmin, and
 	M.E. Shaposhnikov, Phys. Lett. B {\bf 99}, 358 (1981).

 	\bibitem{ABM82}  W.M. Alberico, A. Bottino, and A. Molinari, Phys.
 	Lett. B {\bf 114}, 266 (1982); W.M. Alberico, J. Bernabeu, A.
 	Bottino, and A. Molinari, Nucl. Phys. {\bf A429}, 445 (1984); W.M.
 	Alberico, A. De Pace, and M. Pignone, Nucl. Phys. {\bf A523}, 488
 	(1991).

 	\bibitem{DGM83}  C.B. Dover, A. Gal, and J.M. Richard, Phys. Rev.
 	D {\bf 27}, 1090 (1988); Phys. Rev. C {\bf 31}, 1423 (1985); Nucl.
 	Instr. Meth. A {\bf 284}, 13 (1989).

        \bibitem{HKo98} J. Huefner and B.Z. Kopeliovich, Mod. Phys. Lett.
        A {\bf 13}, 2385 (1998).

 	\bibitem{Bal94}  M. Baldo-Ceolin {\it et al.}, Z. Phys. C {\bf 63},
 	409 (1994).

 	\bibitem{Kam97}  Yu.A. Kamyshkov, in {\it Intersections between
 	Particle and Nuclear Physics}, edited by T.W. Donnelly, AIP Conf.
 	Proc. 412 (AIP, New York, 1997), p. 335.

 	\bibitem{Naz94}  V.I. Nazaruk, Phys. Lett. B {\bf 337}, 328 (1994).

 	\bibitem{Naz98}  V.I. Nazaruk, Phys. Rev. C {\bf 58}, R1884 (1998).

 	\bibitem{DGM95}  C.B. Dover, A. Gal, and J.M. Richard, Phys. Lett.
 	B {\bf 344}, 433 (1995).

        \bibitem{Kri96} M.I. Krivoruchenko, Phys. At. Nucl. {\bf 59}, 1972
        (1996).
 \end{thebibliography}
\end{document}